\documentclass[nofootinbib,notitlepage,superscriptaddress,10pt,aps,pra,twocolumn]{revtex4-1}

\usepackage[utf8x]{inputenc}
\usepackage{amsmath}
\usepackage{amssymb}
\usepackage{graphicx}
\usepackage[normalem]{ulem}
\usepackage{epsf,epsfig}
\usepackage{psfrag,appendix}
\usepackage{color}

\usepackage{braket}

\begin{document}

\title{RELATIVISTIC PLANCK-SCALE POLYMER}

\author{Giovanni AMELINO-CAMELIA}

\author{Michele ARZANO}

\author{$\,\,\,\,\,\,\,\,\,\,\,\,\,\,\,\,\,\,\,\,\,$ Malú Maira DA SILVA}

\author{Daniel H. OROZCO-BORUNDA}
\affiliation{Dipartimento di Fisica, Universit\`a di Roma ``La Sapienza", P.le A. Moro 2, 00185 Roma, Italy}
\affiliation{INFN, Sez.~Roma1, P.le A. Moro 2, 00185 Roma, Italy}

\begin{abstract}
Polymer quantum mechanics has been studied as a simplified picture that reflects
some of the key properties of Loop Quantum Gravity; however, while the fate of relativistic symmetries
in Loop Quantum Gravity is still not established, it is usually assumed that the discrete polymer structure should lead to a breakdown of relativistic symmetries.
We here focus for simplicity on a one-spatial-dimension polymer model and show that relativistic
symmetries are deformed, rather than being broken.
The specific type of deformed relativistic symmetries which we uncover appears to
be closely related to analogous descriptions of relativistic symmetries
in some noncommutative spacetimes.
This also contributes to an ongoing effort attempting to establish whether
the ``quantum-Minkowski limit" of Loop Quantum Gravity is a noncommutative spacetime.
\end{abstract}
\maketitle

\section{INTRODUCTION}
For more than a decade now the study of the fate of relativistic symmetries
in the quantum-gravity realm has been very intense.
The present understanding is based on three possible scenarios:
relativistic symmetries might preserve their ordinary structure, unaffected by Planck-scale effects \cite{rovellispeziale}, or they might be broken by Planck-scale effects, with the emergence
of a preferred frame \cite{gacellis,gambinipullin,urrutia}, or they could be ``Planck-scale deformed",
with a novel role for the Planck scale in the transformation rules among relativistic observers, but still no preferred frame \cite{gacIntJ,gacPhyLett,oldkowalskidsr,magueijosmolin,rovellidsr}.
Frustratingly
the situation remains unclear in Loop Quantum Gravity \cite{rovelli,RovelliLRR,AshLew,Thiem},
one of the most ambitious and rich approaches
to the quantum-gravity problem: because of our current limitations in the analysis of
the Hamiltonian constraint any one of the three mentioned outcomes for the fate of relativistic
symmetries in Loop Quantum Gravity remains possible, with different authors
formulating different intuition for what that outcome might turn out to be.

In light of the complexity of full-fledged quantum-gravity theories, it is being appreciated that
simplified models, capturing some aspects of the more ambitious models, can serve the purpose of both
shading light on the relevant conceptual issues and possibly providing guidance to
Planck-scale phenomenology \cite{gaclrr}. Polymer Quantization (PQ) \cite{shadow}
is relevant from this perspective: it is believed \cite{shadow,frede,bojo,ashbojo,ashpawlow,husain} to provide a simplified model which might capture some aspects of the quantum-gravity realm,
particularly of its Loop-Quantum-Gravity prescription. We are here however concerned with the fact
that it has been claimed \cite{chiou,kajuri} that
PQ should automatically give rise to a preferred frame, whereas,
as mentioned, this is not necessarily the case in Loop Quantum Gravity and Quantum Gravity in general.
Our main objective is to provide evidence in support of the possibility that PQ, perhaps surprisingly in light of
its discrete structure, could be performed without giving rise to a preferred frame. We provide
a definite scenario for the description of Polymer Quantization in terms of Planck-scale-deformed
relativistic symmetries, and interestingly these deformed relativistic symmetries
are rather similar to the ones encountered in the analysis
of some noncommutative spacetimes.
From that perspective our analysis
might also contributes to an ongoing effort \cite{bojopaily,gacmaluronco,FREIDEL,ORITI,scaef,gacfreidkowsmol}
attempting to establish whether
the ``quantum-Minkowski limit" of Loop Quantum Gravity is a noncommutative spacetime.

\section{PRELIMINARIES ON POLYMER QUANTIZATION}
In PQ the Hilbert space realizes a representation of the basic observables of the theory which is unitarily inequivalent to the familiar Schr\"odinger's one \cite{corichi}. It is not just that the state vectors, the inner product and the operators take different form, but rather they also lead to different physical predictions in their domain of applicability.

In order to introduce the polymer representation it is convenient to start from some aspects of the Schr\"odinger representation. We introduce the exponentiated versions of position $q$ and momentum $\pi$ operators, the one-parameter family of unitary operators \cite{shadow,corichi}:
\begin{eqnarray}
U_\mu=e^{i \mu q} \hspace*{1cm} V_\lambda=e^{i \lambda \pi}.
\end{eqnarray}
In  standard quantum mechanics, the operators $U_\mu$ and $V_\lambda$ are well-defined on the Hilbert space of square integrable functions $\mathcal{H}_{Sch}=L^{2}(\mathbb{R},dq)$ and their action on a state  $\psi \in \mathcal{H}_{Sch}$ is given by:
\begin{eqnarray}
U_\mu \psi (q)= e^{i\mu q} \psi (q) \hspace*{1 cm} V_\lambda  \psi (q)= \psi (q+\lambda)\,. \label{basic opertors}
\end{eqnarray}

It is easy to see, using the canonical commutation relation, $[\pi,q]=i$, that $U_\mu$ and $V_\lambda$ satisfy the
following composition rules \cite{shadow}:
\begin{eqnarray}
U_{\mu_1}U_{\mu_2}=e^{i\mu_{1} q}e^{i\mu_2 q}=e^{i(\mu_{1} + \mu_{2}) q}=U_{\mu_1+\mu_2} \, ,
\nonumber
\end{eqnarray}
\begin{eqnarray}
V_{\lambda_1}V_{\lambda_2}=e^{i\lambda_{1} \pi}e^{i\lambda_{2} \pi}=e^{i(\lambda_{1} + \lambda_{2}) \pi}
=V_{\lambda_1 + \lambda_2} \, , \nonumber
\end{eqnarray}
\begin{eqnarray}
U_{\mu}V_{\lambda}=e^{i\mu q}e^{i\lambda \pi}=e^{-i\mu \lambda}e^{i\lambda \pi}e^{i\mu q}
= e^{-i\mu \lambda} V_{\lambda} U_{\mu} \, , \label{algebra de weyl}
\end{eqnarray}
which comprise the Weyl algebra \cite{corichi}. 

According to the Stone-von Neumann theorem \cite{reed} all the representations of the Weyl algebra that fulfill certain conditions of regularity and irreducibility are unitarily equivalent to the Schr\"odinger's one.
In the polymer representation one of the assumptions of the Stone-von Neumann theorem, namely that the operator $V_{\lambda}$ is {\it weakly continuous} in the $\lambda$ parameter, is no longer valid, thus leading to a unitarily inequivalent representation.

The polymer Hilbert space is characterized by a non-countable orthonormal basis $\ket{q}$, labelled by real numbers. The inner product of the Hilbert space is defined as \begin{eqnarray}
\braket{q|q^{\prime}}=\delta_{q,q^{\prime}}. \label{inner product}
\end{eqnarray}
The failure of continuity of $V_\lambda$ in the polymer context is evident if we notice that
\begin{equation}
\lim_{\lambda\rightarrow 0}\braket{q|V_\lambda|q}=\lim_{\lambda\rightarrow 0}\braket{q|q+\lambda}=\lim_{\lambda\rightarrow 0}\delta_{q,q+\lambda}=0 \, .
\end{equation}
It does not matter how small $\lambda$ is, $\ket{q}$ will be always orthogonal to $V_{\lambda}\ket{q}$, while $\braket{q|V_{\lambda=0}|q}=1$. Due to such lack of continuity, the momentum operator $\pi$, seen as the derivative of $V_{\lambda}$ evaluated in $\lambda=0$  is not well defined in the polymer Hilbert space, and thus $V_\lambda$ acquires the status of a fundamental observable. In order to define an operator that plays the role of the momentum in the polymer description, one can get help from the usual description of the operator $V_\lambda$, where $V_\lambda=e^{i \lambda \pi}$. Considering the {\it semiclassical limit} $\lambda \pi << 1$ \cite{corichi,louko}, one can define a momentum operator $P=\frac{1}{\lambda} \sin\left( \lambda \pi \right) \approx \pi +\mathcal{O}(\pi^3)$. Then, in terms of the fundamental observable $V_\lambda$, one has $P=\frac{1}{\lambda}\frac{V_\lambda-V_{-\lambda}}{2i}$.

The polymer Hilbert space is comprised by wavefunctions which take values different from zero in points of the real line which are regularly spaced: $q=q_0+n\lambda$ for a given point $q_0 \in \mathbb{R}$ and $n\in \mathbb{Z}$. Such functions with support on the lattice $\mathcal{L}=\left\lbrace q=q_0+n\lambda, q\in \mathbb{R}  \right\rbrace$, which we denote with $\psi_{q_0}$, belong to a {\it separable} Hilbert space which is a \textit{superselected sector} of the full polymer Hilbert space. States belonging to such a sector cannot be mapped to states in other sectors by any physical operator. Hence a state belonging to the full Hilbert space can be written as a linear superposition of all the functions indexed by $q_0$ belonging to the continuous interval $q_0\in[0, \lambda)$ \cite{capitulo,jd}. This leads to the following characterization of the polymer Hilbert space as a direct sum of superselected sectors $\mathcal{H}_{q_0}$
\begin{eqnarray}
\mathcal{H}_{poly}=\bigoplus_{q_0\in[0, \lambda)}\mathcal{H}_{q_0}\,,
\end{eqnarray}
and thus the most general element of $\mathcal{H}_{poly}$ is not contained in just one superselected Hilbert space $\mathcal{H}_{q_0}$. However, to gain some physical intuition it is customary to restrict the attention to a specific superselected Hilbert space $\mathcal{H}_{q_0}$ and therefore to just work on a fixed regular lattice. We shall also follow this approach, taking for simplicity the point used to fix the lattice as $q_0=0$.

The presence of fundamental discreteness in the polymer framework has led most authors to assume that relativistic symmetries are broken \cite{chiou,kajuri}, with emergence of a preferred frame; however, the study we here report provides evidence in support of the possibility that the polymer framework, contrary to what is commonly expected, can provide the basis for a relativistic picture.

\section{Preliminaries on the manifestly-covariant formulation of Quantum Mechanics}
For our purposes it is natural to work within
the manifestly-covariant formulation of Quantum Mechanics,
in which
one starts from a ``kinematical Hilbert space" where both the time and the spatial coordinates
are self-adjoint operators
\cite{halliwell,gambini,Rovelli}. The Heisenberg algebra of observables is then obtained
 via the imposition of a suitable constraint \cite{Rovelli}, and states that satisfy
 that constraint are said to be in the physical Hilbert space.

Readers will of course find elsewhere more detailed introductions to
the manifestly-covariant formulation of Quantum Mechanics. We shall be here satisfied
with illustrating the logic of this setup by considering a free special-relativistic particle
 in a $(1+1)$-dimensional spacetime. In that case on the kinematical Hilbert space one has
   spacetime coordinates $(q_{t},q)$ and momenta $(\pi_{t},\pi)$ satisfying the canonical commutation relations
\begin{equation}
\begin{split}
[\pi_{t}, \pi]=0,\quad [q_{t},q]=0, \quad [\pi_{t},q_{t}]=-i,
\\
[\pi_{t},q]=0,\quad [\pi,q_{t}]=0,\quad [\pi,q]=i,
\end{split}
\end{equation}
 that are represented on the Hilbert space of square-integrable functions $L^{2}(\mathbb{R}^{2},dq_{t}dq)\sim L^{2}(\mathbb{R}^{2},d\pi_{t}d\pi)$ \cite{AmelinoAstuti}.

States on the physical Hilbert space are those satisfying the Hamiltonian constraint,
 which of course for a free special-relativistic is an on-shell constraint:
\begin{equation}
H\psi=[{\pi_{t}}^{2}-\pi^{2}-m^{2}]\psi=0.
\end{equation}

Basically the kinematical Hilbert space describes abstract points of spacetime (no particles, no physics),
whereas the physical Hilbert space describes
on-shell particles ({\it i.e.} worldlines, rather than points).

\section{Polymer symmetries}

Relevant to the Loop-Quantum-Gravity perspective on PQ is the fact that the usual Schr\"odinger representation
can be seen as an approximate description of PQ via a continuous limit \cite{corichi}.
Here we are interested in the special-relativistic version of PQ, for which, as mentioned, we shall
consider a superselected \cite{shadow} sector of the full polymer Hilbert space. We shall focus
 on
 a free relativistic particle in $(1+1)$ dimensions and \textit{polymerize} only the spatial coordinate \cite{morales} while the temporal coordinate remains continuous. This is the polymer picture usually adopted in  particular in loop quantum cosmology \cite{casting}. The kinematical Hilbert space will be described by the tensor product $\mathcal{H}_{Sch}\otimes \mathcal{H}_{q_0}$.

In order to relate the polymer picture and the Schr\"odinger picture, it is useful to explicitly write the polymer operators of the relativistic system in terms of the operators that characterize the covariant formulation of quantum mechanics \cite{halliwell,gambini,Rovelli,AmelinoAstuti} (time coordinate $q_{t}$, spatial coordinate $q$, time momentum $\pi_{t}$ and spatial momentum $\pi$), to which we
 shall refer as the \textit{pregeometric} representation of the polymer operators
 (in the same spirit of the pregeometric representations in use in
 studies of spacetime noncommutativity \cite{gacmajid,AmelinoAstutiRosati,AmelinoAstuti}).
 We start by introducing, consistently with
 what is frequently done in  the polymer literature \cite{louko},
 \begin{eqnarray}
P=\frac{\sin(\lambda \pi)}{\lambda}  \label{pregeometry},
\end{eqnarray}
where $\lambda$ is a fixed parameter with dimensions of length. In our derivation of
a (deformed-) relativistic description of the polymer picture,
also taking as guidance analogous studies of other quantum spacetimes  \cite{AmelinoAstutiRosati,AmelinoAstuti},
 we combine the translation generator $P$
with a time-translation generator $P_{t}$ and a boost generator $B$ defined as follows:
\begin{eqnarray}
P_t=\pi_t e^{-i\lambda \pi/2}, \hspace*{.7cm} B=e^{-i\lambda \pi}\eta, \label{pregemotrry2}
\end{eqnarray}
where
\begin{equation}\label{pregboost}
\eta=\left(\frac{e^{2i\lambda\pi}-1}{2i\lambda}+i\frac{\lambda}{2}{\pi_{t}}^{2}\right)q_{t}-\pi_{t}q.
\end{equation}
We shall show that this {\it ansatz} for the description of relativistic symmetries
has several properties suggesting that  $P$, $P_{t}$ and $B$ are indeed generators of the (deformed) relativistic symmetries
of the 1-1-dimensional polymer.

We start by verifying that these generators take a state in $\mathcal{H}_{Sch}\otimes \mathcal{H}_{q_{0}}$
and map it into another state still in $\mathcal{H}_{Sch}\otimes \mathcal{H}_{q_{0}}$, {\it i.e.}
\begin{eqnarray}
\psi \in \mathcal{H}_{Sch}\otimes \mathcal{H}_{q_{0}}\rightarrow (S\triangleright \psi) \in \mathcal{H}_{Sch}\otimes \mathcal{H}_{q_{0}}, \label{Symmetrie condition}
\end{eqnarray}
with $S$ any combination of $P$, $P_{t}$ and $B$.
For this purpose we describe a general
 wave function in the polymer picture
in the following way:
\begin{equation}
\int\sum_{j}dq_{t}f(q_{t},j)e^{iq_{t}k_{t}}e^{-iq_{j}k} \, ,
\end{equation}
and therefore by linearity one can focus on the action of the operators on the product of exponentials $$f(q_{t},q_{j})\equiv e^{iq_{t}k_{t}}e^{-iq_{j}k} \, .$$
For the operator  $P$
one finds
\begin{eqnarray}
&& P\triangleright f(q_{t},q_{j})=P\triangleright (e^{iq_{t}k_{t}}e^{-iq_{j}k}) \nonumber \\
&& =\frac{f(q_{t},q_{j}+\lambda)-f(q_{t},q_{j}-\lambda)}{2i\lambda}.
\end{eqnarray}
while for $B$
\begin{eqnarray}
&& B\triangleright f(q_{t},q_{j})=B\triangleright (e^{iq_{t}k_{t}}e^{-iq_{j}k}) \nonumber \\
&& = q_{t}\left(\frac{f(q_{t},q_{j})-f(q_{t},q_{j}-2\lambda)}{2i\lambda}-i\frac{\lambda}{2}(\partial_{0})^{2}f(q_{t},q_{j})\right) \nonumber \\
&& +iq\partial_{0}f(q_{t},q_{j}).
\end{eqnarray}
Therefore the operators $P$ and $B$ satisfy the criteria $(\ref{Symmetrie condition})$,
the action of $P$ and $B$ on a function on $\mathcal{H}_{Sch}\otimes \mathcal{H}_{q_{0}}$
gives a function which is still on $\mathcal{H}_{Sch}\otimes \mathcal{H}_{q_{0}}$.
The same evidently holds also for $P_{t}$.

Let us then notice that the algebra described by $(P_{t},P,B)$,
to which we shall refer as the ``polymer algebra", is characterized by:
\begin{eqnarray}
[P_{t},P]=0, \nonumber
\end{eqnarray}
\begin{eqnarray}
[B,P_{t}]=iP(\sqrt{1-{\lambda}^{2}P^{2}}-i\lambda P)^{1/2}, \nonumber
\end{eqnarray}
\begin{eqnarray}
[B,P]=iP_{t}\sqrt{1-{\lambda}^{2}P^{2}}(\sqrt{1-{\lambda}^{2}P^{2}}-i\lambda P)^{1/2},\label{polyalge}
\end{eqnarray}
so we have a deformed relativistic algebra (a ``DSR-relativistic algebra" \cite{gacIntJ,gacPhyLett}) which recovers the classical Poincar\'e algebra in the limit $\lambda\rightarrow 0$.

Our next task is to show that our candidate algebra of relativistic symmetries
also correctly ``predicts" the Hamiltonian that for independent reasons
has been adopted in the literature for the description of a polymer particle in the Galilean regime. For this purpose we start by noticing
that the Casimir of the polymer algebra is given by
\begin{equation}\label{casimir}
C=\frac{2}{{\lambda}^{2}}(1-\sqrt{1-{\lambda}^{2}P^{2}})-{P_{t}}^{2} \, ,
\end{equation}
which in terms of $\pi$ takes the form
\begin{equation}\label{polymercasimir}
C=\frac{4}{{\lambda}^{2}}\sin^{2}\left(\frac{{\lambda} \pi}{2}\right)-{P_{t}}^{2} \, .
\end{equation}
In preparation for considering the Galilean regime we  make the replacements
\begin{equation}
C\rightarrow -m^{2}c^{2}, \quad P_{t}\rightarrow \frac{E}{c} \, ,
\end{equation}
$m$ being the mass and $c$ the speed of light, so that (\ref{polymercasimir})
takes the form
\begin{equation}
E=mc^{2}\sqrt{1+\frac{4}{{\lambda}^{2}m^{2}c^{2}}\sin^{2}\left(\frac{\lambda \pi}{2}\right)}.
\end{equation}
Thus, in the Galilean regime ($c\rightarrow \infty$) one has
\begin{equation}\label{galileanlimit}
E=\frac{1-\cos(\lambda \pi)}{{\lambda}^{2}m} \, ,
\end{equation}
where of course we dropped the constant contribution to energy, $mc^{2}$, which is unnoticeable
in the Galileian regime (since there is no particle production).
Eq. (\ref{galileanlimit}) is indeed exactly the Hamiltonian that is often used in
the literature \cite{corichi,louko} in the description of a free Galilean-regime polymer particle.

\section{A possible connection with the $\kappa$-Poincar\'e algebra}
We have provided evidence in support of the thesis that PQ, rather than breaking relativistic
symmetries as usually assumed \cite{chiou,kajuri}, may be characterized by a deformation
of relativistic symmetries (without a preferred frame).
Since the polymer picture is being considered as a simplified version of Loop Quantum Gravity,
our results strengthen the case (already suggested by several authors \cite{bojopaily,gacmaluronco,FREIDEL,ORITI,scaef,gacfreidkowsmol})
for the emergence of DSR-relativistic symmetries in the ``Minkowski regime" of Loop Quantum Gravity.
Our next objective is to consider a possible connection between our results
and the most studied possibility for the description
of these Minkowski-regime relativistic symmetries, which involves the
 $\kappa$-Poincar\'e Hopf algebra \cite{bojopaily,gacmaluronco,FREIDEL,ORITI,scaef,gacfreidkowsmol}.

 It had already been observed more than 20 years ago \cite{celeghini} that phonons in certain condensed matter systems,
 with atomic-structure discreteness, are governed by equations
 of motion that are $\kappa$-Poincar\'e
 invariant. On the technical side some evidence for a role of discreteness
 in the structure of $\kappa$-Poincar\'e has been uncovered so far only
in studies of an associated differential calculus \cite{majiddiff,gacmajid,robert}.
We here contribute to these investigations by observing that our polymer algebra is
related to the  $\kappa$-Poincar\'e algebra in a rather simple manner.

Let us consider, within our polymer algebra, the following relationships
between the operators $P_{t},P,B$, generators of our polymer algebra,
and some other operators $\mathcal{P}_{t},\mathcal{P},\mathcal{B}$:
\begin{eqnarray}
P_{t}=\mathcal{P}_{t}e^{i\lambda\mathcal{P}/2}, \nonumber
\end{eqnarray}
\begin{eqnarray}\label{changebasis}
P=\frac{\sin(\lambda\mathcal{P})}{\lambda},\quad B=\mathcal{B} \, .
\end{eqnarray}
It is straightforward to verify that upon these replacements the commutators of
our polymer algebra take the form
\begin{eqnarray}
[\mathcal{P}_{t},\mathcal{P}]=0, \quad [\mathcal{B},\mathcal{P}]=i\mathcal{P}_{t}, \nonumber
\end{eqnarray}
\begin{eqnarray}\label{bicross}
[\mathcal{B},\mathcal{P}_{t}]
=\frac{1-e^{-2i\lambda\mathcal{P}}}{2\lambda}+\frac{\lambda}{2}{P_{t}}^{2} \, ,
\end{eqnarray}
and these are exactly the commutators that characterize \cite{majRue,lukRueg}
 the $\kappa$-Poincar\'e algebra, if we exchange the roles of $\mathcal{P}_{t}$ and $\mathcal{P}$
 (and upon of course reinterpreting the polymer parameter $\lambda$ in terms
 of the parameter $\kappa$ of the $\kappa$-Poincar\'e algebra).

\section{OUTLOOK}
We here provided results in support of a rather unexpected scenario
for the quantum-gravity realm, in which spacetime discretization and
relativistic covariance (however deformed) coexist.
As already stressed above we feel that our results are particularly intriguing since 
PQ is viewed as a toy model for Loop Quantum Gravity and our results may contribute
both to the debate of the fate of relativistic symmetries in Loop Quantum Gravity
and to the investigations of the possibility that the ``quasi-Minkowski regime"
of Loop Quantum Gravity might be described in terms of spacetime noncommutativity.

Some  further checks are needed in order to reach a fully established picture: we verified a particular significant subset
of expected properties of a relativistic theory, but surely more tests
would be needed in order to be confident of the overall consistency
of the scenario. Among these we feel particular interest is deserved
by the introduction of interactions among particles, to be  handled
consistently within a quantum-field-theory formulation of the polymer.

While we might be near to establishing this picture
for the 2D polymer toy model (with only the spatial coordinate polymerized),
 of course ultimately we would want to see all  this at work
 in more realistic models. We expect that already  the generalization
 to a 4D polymer structure should be rather nontrivial.

\section*{ACKNOWLEDGMENTS} 
We would like to thank Danilo Latini for carefully reading the manuscript and for providing valuable feedback. M. M. S. thanks the Brazilian Federal Agency for Support and Evaluation of Graduate  Education - CAPES (within the Ministry of Education of Brazil) for its financial support. D.H.O.B acknowledges financial   support from CONACyT grant No.409825 and previous support through the CONACyT grant 237351-“Implicaciones f\'isicas de la estructura del espacio tiempo".


\begin{thebibliography}{}
\bibitem{rovellispeziale} C. Rovelli, S. Speziale,  Phys. Rev. \textbf{D83}, 104029  (2011).

\bibitem{gacellis} G. Amelino-Camelia, J.Ellis, N. E. Mavromatos, D. V. Nanopoulos, S. Sarkar. Nature, 393(6687), 763-765 (1998).

\bibitem{gambinipullin} R. Gambini, J.  Pullin. gr-qc/0110054 (2001).

\bibitem{urrutia} J. Alfaro, H. A. Morales-Técotl, M. Reyes, L. F. Urrutia. Phys. Rev. \textbf{D70}, 084002 (2004).

\bibitem{gacIntJ} G. Amelino-Camelia, Int. J. Mod. Phys. \textbf{D11}, 35 (2002).

\bibitem{gacPhyLett} G. Amelino-Camelia, Phys. Lett. \textbf{B510}, 255 (2001).


\bibitem{oldkowalskidsr} J. Kowalski-Glikman, gr-qc/0603022 (2006).

\bibitem{magueijosmolin} Phys. Rev. \textbf{D67}, 044017 (2003).

\bibitem{rovellidsr} C. Rovelli.  gr-qc/0808.3505 (2008).



\bibitem{rovelli} C. Rovelli. Quantum gravity. Cambridge university press (2007).

\bibitem{RovelliLRR} C. Rovelli, Living Rev. Rel.  1, 1
            (1998).

\bibitem{AshLew} A. Ashtekar and J. Lewandowski, Class. Quant. Grav. R53, 21 (2004).

\bibitem{Thiem} T. Thiemann, Lect. Notes Phys. \textbf{631}, 41 (2003).

\bibitem{gaclrr} G. Amelino-Camelia, Living Rev. Rel. 16 (2013) 5.

\bibitem{shadow} A. Ashtekar, S. Fairhurst, J. Willis, Class. Quant.Grav., 20  1031-1062 (2003).

\bibitem{frede} K. Fredenhagen, F. Reszewski, Class. Quant. Grav. 23 6577 (2006).

\bibitem{bojo} M. Bojowald, Living Rev. Rel. 8, 11 (2005).

\bibitem{ashbojo} A. Ashtekar, M. Bojowald, J. Lewandowski, Adv. Theor. Math. Phys. 7 233 (2003).

\bibitem{ashpawlow} A. Ashtekar, T. Pawlowski, P. Singh, Phys. Rev. \textbf{D74} 084003 (2006).

\bibitem{husain} V. Husain and O. Winkler, Phys. Rev. \textbf{D75} 024014 (2007).

\bibitem{chiou} D. Chiou, Class. Quant. Grav., 24 (2007) 2603-2620.

\bibitem{kajuri} G. Date, N. Kajuri, Class. Quantum Grav. 30 (2013) 075010.

\bibitem{bojopaily} M. Bojowald, G.M. Paily, Phys. Rev. \textbf{D87}, 044044 (2013) 4.

\bibitem{gacmaluronco} G. Amelino-Camelia, M. M. da Silva, M. Ronco, L. Cesarini, O. M. Lecian, Phys. Rev. \textbf{D95}, 024028 (2017).

\bibitem{FREIDEL} L. Freidel, Etera R. Livine, Phys. Rev. Lett. \textbf{96}, 221301 (2006).

\bibitem{ORITI} D. Oriti, T. Tlas, Phys. Rev. \textbf{D74}, 104021 (2006).

\bibitem{scaef} B.E. Schaefer, Phys. Rev. Lett. \textbf{82}, 4964 (1999).

\bibitem{gacfreidkowsmol} G. Amelino-Camelia, L. Freidel, J. Kowalski-Glikman, L. Smolin, Phys. Rev. \textbf{D84}, 084010 (2011).



\bibitem{corichi} A. Corichi, T. Vukasinac, J. A. Zapata, Phys. Rev. \textbf{D76}, 044016 (2007).

\bibitem{reed} M .Reed, B. Simon, Methods of modern mathematical physics. vol. 1. Functional analysis. Academic (1980).

\bibitem{louko} G. Kunstatter, J. Louko, J. Ziprick, Phys. Rev. \textbf{A79}, 032104 (2009).

\bibitem{capitulo} Henri Poincar\'e y David Hilbert: los \'ultimos universalistas y los fundamentos de la fisica matem\'atica moderna, Universidad Aut\'onoma Metropolitana (2013).

\bibitem{jd} E. Flores-González, H.A. Morales-Técotl, J.D. Reyes, Annals of Physics, 336 (2013), 394-412.

\bibitem{halliwell} J. J. Halliwell, Phys. Rev. \textbf{D64}, 044008 (2001).

\bibitem{gambini} R. Gambini, R. A. Porto, Phys. Rev. \textbf{D63}, (2001).

\bibitem{Rovelli} M. Reisenberger, C. Rovelli, Phys. Rev. \textbf{D65}, 125016 (2002).

\bibitem{AmelinoAstuti} G. Amelino-Camelia, V. Astuti, Int. J. Mod. Phys. \textbf{D24} (2015) n.10, 1550073.

\bibitem{morales} H. A. Morales-Técotl, S. Rastgoo, J. C. Ruelas, Phys. Rev. \textbf{D95}, 065026 (2017).

\bibitem{casting} A. Ashtekar, M. Campiglia, A. Henderson, Class. Quant. Grav. 27 (2010) 135020.

\bibitem{gacmajid} G. Amelino-Camelia, S. Majid, Int. J. Mod. Phys \textbf{A15} (2000) 4301-4324.

\bibitem{AmelinoAstutiRosati} G. Amelino-Camelia, V. Astuti, G. Rosati, Eur. Phys. J. C 73 (2013) 2521.

\bibitem{celeghini} F. Bonechi, E. Celeghini, R. Giachetti, E. Sorace, M. Tarlini, Phys. Rev. Lett. \textbf{68} (1992) 3718-3720.

\bibitem{majiddiff} S. Majid, J. Math. Phys. 46 (2005) 103520.

\bibitem{robert} R. Oeckl, J. Math. Phys. 40 (1999) 3588-3603.

\bibitem{majRue} S. Majid, H. Ruegg, Phys. Lett. \textbf{B334}, 348 (1994).

\bibitem{lukRueg} J. Lukierski, H. Ruegg, W.J. Zakrzewski, Ann. Phys. \textbf{243}, (1995) 90.





%
%




\end{thebibliography}
\end{document}